\documentclass[review]{elsarticle}

\usepackage{lineno}

\usepackage{multirow,setspace,amssymb,amsmath,graphicx,color,rotating,subfigure}
\usepackage{lineno}
\usepackage{textcomp}
\usepackage{booktabs}    
\usepackage{CJK}
\usepackage{bm}%
\usepackage{rotating}
\usepackage{diagbox}
\usepackage{epsfig}
\usepackage{dcolumn}
\usepackage{epstopdf}
\usepackage{natbib}
\biboptions{sort&compress}
\usepackage[hidelinks]{hyperref}

\usepackage[title]{appendix}%
\newcommand{\customcup}[2]
{\bigcup\limits_{#1}^{#2}}

\usepackage[ruled,linesnumbered]{algorithm2e}

\begin{document}

\begin{frontmatter}

\title{Influence maximization in multilayer networks based on adaptive coupling degree}

\author[inst1]{Su-Su Zhang}
\author[inst1]{Ming Xie}
\author[inst1]{Chuang Liu\corref{ca1}}\ead{liuchuang@hznu.edu.cn}
\author[inst1,inst2]{Xiu-Xiu Zhan\corref{ca1}}\ead{zhanxiuxiu@hznu.edu.cn}

\address[inst1]{Alibaba Research Center for Complexity Sciences, Hangzhou Normal University, Hangzhou, 311121, P. R. China}
\address[inst2]{College of Media and International Culture, Zhejiang University, Hangzhou 310058, PR China}

\cortext[ca1]{Corresponding authors.}

\begin{abstract}
Influence Maximization(IM) aims to identify highly influential nodes to maximize influence spread in a network. Previous research on the IM problem has mainly concentrated on single-layer networks, disregarding the comprehension of the coupling structure that is inherent in multilayer networks. To solve the IM problem in multilayer networks, we first propose an independent cascade model (MIC) in a multilayer network where propagation occurs simultaneously across different layers. Consequently, a heuristic algorithm, i.e., Adaptive Coupling Degree (ACD), which selects seed nodes with high spread influence and a low degree of overlap of influence, is proposed to identify seed nodes for IM in a multilayer network. By conducting experiments based on MIC, we have demonstrated that our proposed method is superior to the baselines in terms of influence spread and time cost in 6 synthetic and 4 real-world multilayer networks.

\end{abstract}


\end{frontmatter}

\section{Introduction}
\label{Intro}
\makeatletter
\newcommand{\rmnum}[1]{\romannumeral #1}
\newcommand{\Rmnum}[1]{\expandafter\@slowromancap\romannumeral #1@}
\makeatother

The objective of the classical influence maximization (IM) problem is to strategically pick a limited number of nodes in order to maximize the propagation of influence in a network based on a particular spreading model~\cite{kempe2003maximizing}. Solving the problem of IM can be advantageous for a variety of purposes, including improving product promotion~\cite{chen2010scalable,domingos2001mining}, limiting the spread of contagious disease~\cite{cheng2020outbreak}, and augmenting the dissemination of information on a social media platform~\cite{zhang2014recent}. To address more realistic issues, the traditional IM problem has been extended to a variety of practical situations by introducing additional restrictions~\cite{leskovec2007cost,goyal2011celf++,xie2023efficient,xie2023vital,yang2023new,li2023influence}. For example, the budgeted IM problem~\cite{nguyen2013budgeted,banerjee2019combim} considers a limited budget to select seeds to maximize influence. The competitive IM problem~\cite{bharathi2007competitive,bozorgi2017community} assumes that there is more than one piece of information spread in a network, and thus algorithms are proposed for the IM problem under a competitive scenario. Furthermore, the targeted IM problem~\cite{cai2020target,calio2021attribute,liang2023targeted} aims to find seeds to maximize influence on a targeted set of users.

Most algorithms created to address the IM problem are intended for single-layer networks, which only consider information spread through one social platform. However, the reality is that a user can register on more than one social platform and have various connections on different platforms~\cite{kivela2014multilayer,watts1998collective}. Therefore, information may be transmitted through various platforms, and we need to consider the IM problem under the multilayer network scenario, which is used to represent connections between users on various social platforms~\cite{buccafurri2014model,al2016identifying,kuzmanov2013protein,kenley2011detecting,gallotti2015multilayer}.  Recently, many researchers have proposed different algorithms to solve the IM problem in multilayer networks, including greedy-based algorithms~\cite{kuhnle2018multiplex}, heuristic algorithms~\cite{katukuri2022cim,rao2022cbim} and meta-heuristic algorithms~\cite{hu2023nodes,wang2019finding,li2023influence-csf}. For example, Kuhnle et al. proposed a KSN algorithm that selects seed nodes from each layer greedily~\cite{kuhnle2018multiplex}. Although KSN can find the optimal seeds, the complexity of KSN limits its capacity to be scaled to multilayer networks with large sizes.  Consequently, Katukuri et al. designed CIM which is based on maximal cliques to efficiently find seeds for IM~\cite{katukuri2022cim}. However, CIM is suitable for networks with clique structures, which makes it impossible to apply to most of the real-world multilayer networks that are sparse in nature. Furthermore, Rao et al. further proposed CBIM~\cite{rao2022cbim} which uses a community detection algorithm to partition a multilayer network into communities and calculates the edge weight sum (EWS) for nodes within each community to identify the most influential nodes. 
  
Despite the efforts that have been made on the IM problem in multilayer networks, most researchers assume that information spreads independently on each layer, and thus the algorithms are designed to select seeds separately in different layers. That is to say, majority of the algorithms designed to find seeds in multilayer networks cannot solve the IM problem in reality, where information generally diffuses across layers. In view of this, we first propose a spreading model that allows information to spread across layers and propose a heuristic algorithm to solve the IM problem in multilayer networks.

For this purpose, our research has yielded the following contributions. First, we propose a heuristic algorithm named Adaptive Coupling Degree (ACD) for influence maximization in multilayer networks. ACD comprehensively evaluates nodes considering neighbor information across layers and then iteratively selects seed nodes with high spread influence and low overlap of influence with each other. Second, we design an independent cascade model in a multilayer network (MIC) to quantify the influence spread of different algorithms. Last but not least, experiments demonstrate that our proposed algorithm outperforms baseline algorithms in synthetic and real-world multilayer networks.

\section{Preliminary definition}
\label{sec:Definition}
\subsection{Influence Maximization in multilayer networks}

We denote an undirected and unweighted multilayer network as $G = \{G_1, G_2, \\ \cdots, G_L\}$, where layer $l$ of $G$ is given by $G_l = (V, E_l)$. Therefore, in every layer of $G$, we have the same node set, which is denoted as $V = \{v_1, v_2, \cdots, v_N\}$, but different edge sets. The set of edges in $G_l$ is represented as $E_l = \{e_1, e_2, \cdots, e_h\}$, where $h$ represents the number of edges in $G_l$.
The aim of Influence Maximization (IM) in a multilayer network is to find $K$ seed nodes that will cause the most spread of influence throughout the network under a particular spreading model. This can be expressed mathematically as:

\begin{equation}\label{equ:PixelColorContrast}
    \mathrm{argmax} \{|\customcup{\text{$l$=1}}{\text{L}}V^{l}_a(S)|,|S| = K , S \subseteq V\},
\end{equation}
where $V^{l}_a(S)$ is the set of active nodes in layer $G_l$, $S$ represents the set of seed nodes,  and $K$ is the size of $S$.

\subsection{Independent cascade model in a multilayer network}\label{Propagation Model}
We propose an independent cascade model that is suitable to replicate the dynamics of propagation in a multilayer network (MIC), where propagation occurs simultaneously across different layers. In MIC, we classify nodes into two different categories, that is, active nodes and inactive nodes. Each node can infect its neighbors only once. When a node $v_i$ is infected at one layer of the network and turns to the active state, it will be active across all layers and will become infectious and can infect its neighbors at any layer. The details of MIC are given below.

\begin{itemize}
\item Initially, the seeds are assigned to the active state, and the remaining nodes are in the inactive state.

\item  At time step $t$, assuming that the contagion process reaches layer $l$, we first find the active nodes in this layer. For each active node $v_i$,  the inactive neighbors of $v_i$ in layer $l$ are denoted as $N^l_{ina}(v_i)$. For each $v_j \in N^l_{ina}(v_i)$, it will be activated by $v_i$ with independent probability $p$ and will be in an active state if activated.

\item At time step $t+1$, the contagion process will occur in layer $l+1$ and performs similarly to those at time step $t$. It should be noted that if $l+1 > L$, the propagation at $t+1$ will be in the first layer.

\item The contagion process continues until no new nodes are activated.
\end{itemize}

\begin{figure*}[!ht]
\centering
	\includegraphics[width=\linewidth]{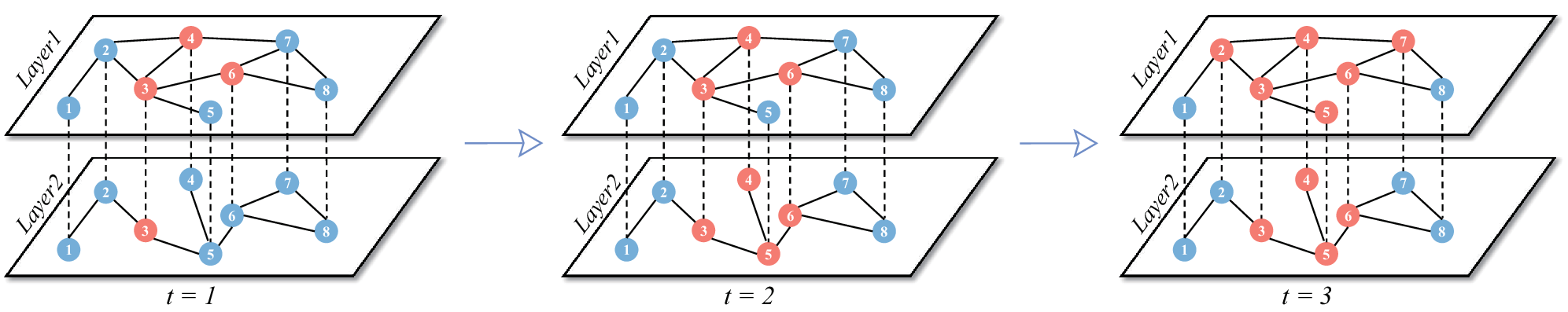}
 \caption{A schematic diagram of the MIC spreading dynamics in a two-layered network.}
 \label{fig:multi-spread}
\end{figure*}

We show an example of MIC in a two-layered network in Figure~\ref{fig:multi-spread}. We assign a seed node as $v_3$. At time step $t=1$, the inactive neighbors of node $v_3$ at the first layer are given by the set $N^1_{ina}(v_3) = \{v_2, v_4, v_5, v_6\}$ and each will be activated by $v_3$ with probability $p$. As shown in the figure, nodes $v_4$ and $v_6$ are successfully activated. Subsequently, the activated nodes at the beginning of step $t=2$ are $v_3$, $v_4$, and $v_6$, and will activate each of their inactive neighbors with probability $p$ in the second layer. As a consequence, node $v_5$ is successfully activated. At step $t=3$, the propagation will return to layer 1 and the newly activated nodes are $v_2$ and $v_7$.

\subsection{Data description}\label{Data description}
In this section, we give a summary of four multilayer networks\footnote{\url{https://manliodedomenico.com/data.php}} that exist in the real world, focusing on their fundamental topological characteristics.
Detailed descriptions of the datasets are given below:

\textbf{CKM Physicians Innovation Network (CKM)~\cite{coleman1957diffusion}}. CKM dataset illustrates how physicians used the new drug tetracycline in four towns. The interactions of each layer between physicians are constructed according to the following three questions: \romannumeral 1) If you need advice about therapy, who do you usually ask? \romannumeral 2) Which three or four physicians do you usually talk to about cases or treatments during a typical week, such as last week? \romannumeral 3)  Can you tell me the first names of the three people you spend the most time with socially? The three layers are named shortly as Adv, Dis and Fri, respectively.
We construct three two-layered networks using this data: Adv-Dis, Adv-Fri, and Dis-Fri.

\textbf{London Multiplex Transport Network (Transport)~\cite{de2014navigability}}. This dataset was obtained from the official website of the London Transport Authority in 2013. It includes three separate layers: the underground (U), the overground (O), and the Docklands light railway (D) network. The nodes represent the transport stations in London, while the edges represent the existing routes between the stations while using different transportation tools. We construct three two-layer networks based on this dataset, i.e., U-O, U-D, and O-D.

\textbf{Arxiv Netscience Multiplex (Arxiv)~\cite{de2015identifying}}. The Arxiv dataset includes co-authorship between authors in three distinct Arxiv categories: mathematical physics (MP), quantitative biology (QB), and quantitative biomolecules (QBM). In each layer of the network, node represents the authors and the edges represent the co-authorship between the authors. Consequently, we can construct three two-layered networks using this dataset, i.e., MP-QBM, MP-QB, and QBM-QB. 

\textbf{C.elegans Multiplex Gpi Network (C.elegans)~\cite{de2015structural}}. The C.elegans dataset indicates the genetic-protein interaction (GPI) of C.elegans. Each layer of C.elegans represents one type of interaction within the organism, namely physical interactions (Phy), suppressive interactions (SI) or additive interactions (AI). In C.elegans, nodes correspond to proteins, while edges encode protein-protein interaction relationships. Based on this dataset, we construct three two-layered networks: Phy-AG, Phy-SG, and SG-AG.

Using the four multiplexes mentioned above, we can obtain $12$ two-layer networks, which are subsequently used to evaluate the performance of the algorithms to solve the IM problem. A comprehensive summary of the basic topological properties of the networks is presented in Table~\ref{topological properties}.

\begin{table}[htbp]
	\centering
	\caption{\label{topological properties} \small Topology properties of empirical two-layered networks. $N$ and $h$  represent the number of nodes and edges in each layer, respectively. $\langle k \rangle$ denotes the average degree of the nodes, $C$ represents the clustering coefficient and $\rho$ represents the link density in each layer.}

	\label{tab:1} 
    \resizebox{0.65\linewidth}{!}{
	\begin{tabular}{cccccccc}
		\hline\hline\noalign{\smallskip}	
		Dataset& Two-layer &Layer & $N$ & $h$ &$\langle k \rangle$& $C$ & $\rho$ \\
		\noalign{\smallskip}\hline\noalign{\smallskip}

        &\multirow{2}{*}{Adv-Dis}&Adv&234 &449 &3.8376 &0.2600 &0.0165\\
        & &Dis &234 &498 &4.2600 &0.2598 &0.0183 \\
        \cmidrule(r){2-8}
        \multirow{2}{*}{CKM}&\multirow{2}{*}{Adv-Fri}&Adv&238 &449 &3.7731 &0.2600 &0.0159\\
        & &Fri &238 &423 &3.5500 &0.2108 &0.0150\\
        \cmidrule(r){2-8}
        &\multirow{2}{*}{Dis-Fri}&Dis &241 &498 &4.1328 &0.2598 &0.0172\\
        & &Fri &241 &423 &3.5100 &0.2108 &0.0146 \\
		\noalign{\smallskip}\hline\noalign{\smallskip}
        &\multirow{2}{*}{U-O}&U &330 &312 &1.8909 &0.0311 &0.0057\\
        &&O &330 &83 &0.5000 &0.0000 &0.0015 \\
        \cmidrule(r){2-8}
        \multirow{2}{*}{Transport}&\multirow{2}{*}{U-D}&U &311 &312 &2.0064 &0.0311 &0.0065\\
        &&D &311 &46 &0.3000 &0.0185 &0.0010\\
        \cmidrule(r){2-8}
        &\multirow{2}{*}{O-D}&O &126 &83 &1.3175 &0.0000 &0.0105\\
        &&D &126 &46 &0.7300 &0.0185 &0.0058 \\
         \noalign{\smallskip}\hline\noalign{\smallskip}
        &\multirow{2}{*}{MP-QBM}&MP &2152 &592 &0.5502 &0.8579 &0.0003\\
        &&QBM &2152 &4423 &4.1100 &0.7670 &0.0019 \\
        \cmidrule(r){2-8}
        \multirow{2}{*}{Arxiv}&\multirow{2}{*}{MP-QB}&MP&1013 &592 &1.1688 &0.8579 &0.0012\\
        &&QB &1013 &868 &1.7100 &0.6908 &0.0017\\
        \cmidrule(r){2-8}
        &\multirow{2}{*}{QBM-QB}&QBM &2510 &4423 &3.5243 &0.7670 &0.0014\\
        &&QB &2510 &868 &0.6900 &0.6908 &0.0003 \\
        \noalign{\smallskip}\hline\noalign{\smallskip}
        &\multirow{2}{*}{Phy-AI}&Phy &1224 &270 &0.4412 &0.1743 &0.0004\\
        &&AI &1224 &2115 &3.4600 &0.1338 &0.0028\\
        \cmidrule(r){2-8}
        \multirow{2}{*}{Celegans}&\multirow{2}{*}{Phy-SI}&Phy&333 &270 &1.6216 &0.1743 &0.0049\\
        &&SI &333 &160 &0.9600 &0.0162 &0.0029\\
        \cmidrule(r){2-8}
        &\multirow{2}{*}{AI-SI}&AI &1121 &2115 &3.7734 &0.1338 &0.0034\\
        &&SI &1121 &160 &0.2900 &0.0162 &0.0003 \\
        \noalign{\smallskip}\hline
	\end{tabular}
    }
\end{table}

\newpage
\clearpage
\section{Algorithms}\label{Algorithms}
\subsection{Adaptive Coupling Degree}
\begin{figure*}[!t]
\centering
	\includegraphics[width=\linewidth]{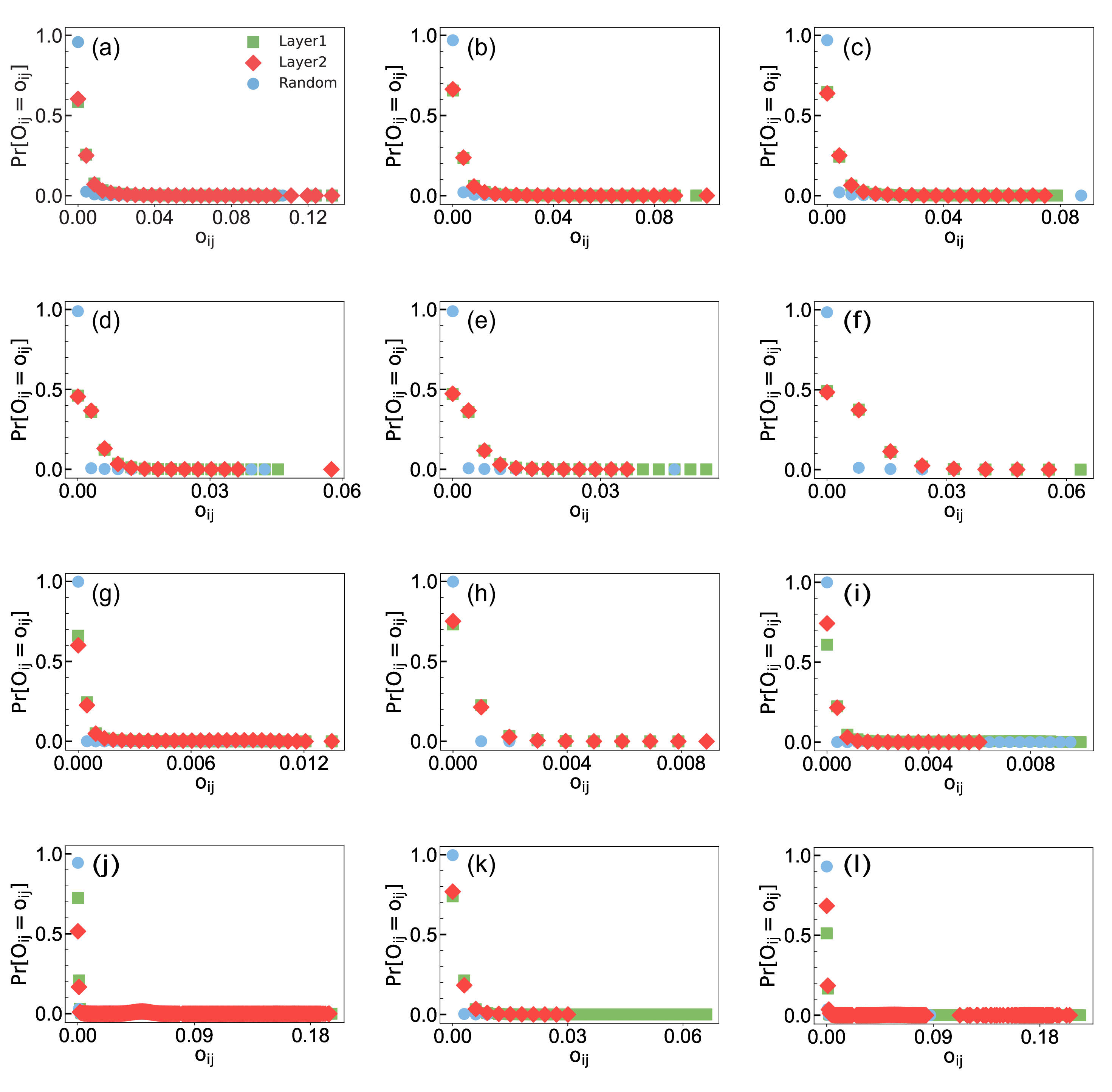}
 \caption{Influence overlap distribution of the pairs of neighboring nodes in the first layer (green square), the second layer (red diamond), and pairs of randomly selected nodes (blue circle). We show the results for the following two-layered networks: (a) Adv-Dis; (b) Adv-Fri; (c) Dis-Fri; (d) U-O; (e) U-D;(f) O-D;(g) MP-QBM; (h) MP-QB; (i) QBM-QB; (j) Phy-AI; (k) Phy-SI; (l) SI-AI.}
 \label{fig: overlap}
\end{figure*}

To solve the influence maximization problem, we not only need to choose nodes with a high influence but also need to consider influence overlap between nodes. If a node $v_i$ with high influence is chosen as the seed, the nodes that have a large influence overlap with $v_i$ should have a low chance of being chosen as the seeds, as they cannot help increase the influence margin. We demonstrate the distribution of the influence overlap of nodes in empirical multilayer networks in Figure~\ref{fig: overlap}. Particularly, 
the influence overlap $o_{ij}$ between nodes $v_i$ and $v_j$ is denoted as $o_{ij}=\frac{|I(v_i)\cap I(v_j)|}{N}$, where $I(v_i)$ denotes the spreading influence of node $v_i$ when $v_i$ is chosen as the seed. We compare the distribution of the influence overlap of the pairs of neighboring nodes in the first layer, in the second layer, and the pairs of randomly selected nodes. The results show that, compared to the random case, the influence overlap distribution of the neighboring nodes is more oriented to the right, indicating that neighboring nodes have a much larger influence overlap.

\newpage
\clearpage
Motivated by the above analysis, we propose an adaptive coupling degree (ACD) algorithm for seed node selection in a multilayer network, which aims to find nodes with high spread influence and low influence overlap. To start, we define the coupling degree $D\_c(v)$ of a node $v$ as the number of non-repetitive neighbors across $L$ layers, the formula can be described as: 

\begin{equation}\label{equ:PixelColorContrast}
    D\_c(v)=|N_1(v)\cup N_2(v)\cup\dots\cup N_L(v)|,
\end{equation}
where $N_l(v) (l=1, 2, \cdots, L)$ represents the neighboring set of $v$ in the $l$-th layer. Then we iteratively compute the adaptive coupling degree of each node and choose the nodes with the highest adaptive coupling degree as the seeds. The initialization of the adaptive coupling degree of each node is given by the value of its coupling degree, i.e., $D\_c^0 = (D\_c^0(v_1), D\_c^0(v_2),\cdots, D\_c^0(v_N))=(D\_c(v_1), D\_c(v_2),\cdots, D\_c(v_N))$. We further define the number of seed nodes in the neighborhood of node $v$ as $Q(v)$. Therefore, the initial number of seed nodes in the neighborhood of every node is 0, and thus we obtain a vector $Q^0= (Q^0(v_1), Q^0(v_2),\cdots, Q^0(v_N))=(0, 0, \cdots,0)$. Therefore, the procedures to iteratively select the seeds are given below.

\begin{itemize}
\item At the first step, node $v_i$ with the highest adaptive coupling degree ($D\_c^0(v_i) = \max\{D\_c^0\}$) is chosen as the first seed node and added to the seed set $S$. Then, we compute the number of seed nodes in the neighborhood of every node and get $Q^1= (Q^1(v_1), Q^1(v_2),\cdots, Q^1(v_N))$. The adaptive coupling degree of node $v_q$ is updated using the following equation:
\begin{equation}\label{equ:1}
    D\_c^1(v_q) = D\_c^{0}(v_q) - e^{Q^{1}(v_q)}.
\end{equation}

\item At time step $t$, we order the nodes according to their adaptive coupling degree, which is determined by the vector $D\_c^{t-1}$, and pick the one with the highest degree from $V\setminus S$ and add it to $S$. Similarly, we update the vector $Q^{t-1}$ based on the number of seed nodes in the neighbors of each node and obtain $Q^t$. Therefore, the adaptive coupling degree of every node is $v_q$ updated based on $D\_c^{t-1}(v_q)$ and $Q^t(v_q)$:
\begin{equation}\label{equ:2}
    D\_c^t(v_q) = D\_c^{t-1}(v_q) - e^{Q^{t}(v_q)}.
\end{equation}

\item The algorithm terminates when $K$ seeds are selected.
\end{itemize}

In ACD, we use the adaptive coupling degree to quantify a node's spread influence, the nonlinear penalty factor introduced by $Q$ is used to filter nodes with low overlap influence, as shown in Eq.~\ref{equ:1} and~\ref{equ:2}.
The pseudocode of ACD is provided in Algorithm~\ref{ACD}.
\begin{algorithm*}[htp]
\caption{\label{ACD}Adaptive Coupling Degree (ACD)}\label{algorithm}
  \SetKwData{Left}{left}\SetKwData{This}{this}\SetKwData{Up}{up}
  \SetKwFunction{Coupling}{Coupling}\SetKwFunction{FindCompress}{FindCompress}
  \SetKwInOut{Input}{Input}\SetKwInOut{Output}{Output}
 
  \Input{Size of seed nodes $K$\\
  A multilayer network $G = \{G_1, G_2, \cdots, G_L\}$}
  \Output{Seed node set $S$}

$\textbf{Initialization :} S = \emptyset , Q^0 \xleftarrow{} 0, D\_c^0 \xleftarrow{}$ Coupling degree of each node\\

    \While{$|S|\leq K$}{
        $t \xleftarrow{} |S| $\\
        $v_{i}  (v_{i} \in V \backslash S) \xleftarrow{} \max\{D\_c^{t}$\}\\
        $S = S \cup \{ v_{i} \}$\\
        $Q^{t+1} = Q^{t}$ \\
        \For{neighbors ${v_j}$ of $v_{i}$ (${v_j} \in V  \backslash S$)}{
        $Q^{t+1}(v_j) = Q^{t}(v_j) + 1 $\;}
    
    
        \For{$v_q$ in $V \backslash S$}{
        $D\_c^{t+1}(v_q) = D\_c^{t}(v_q) - e^{Q^{t+1}(v_q)}$\\
        }
    }
\end{algorithm*}

\subsection{Baselines}

To assess the effectiveness of our algorithm, we introduce three state-of-the-art previously proposed methods and extend five centrality-based methods as baselines. We show the details of each of them below.

\textbf{CBIM}~\cite{rao2022cbim} is a community-based influence maximization algorithm that considers both node degree and the distance between the nodes. CBIM first conducts community detection in each layer by computing the similarity between nodes. Concretely, the similarity between node $v_u$ and $v_q$ in layer $l$ is $DSC_l(v_u,v_q) = \dfrac{2*|N_l(v_u)\cap N_l(v_q)|}{|N_l(v_u)| + |N_l(v_q)|}$, where $N_l(v_u)$ and $N_l(v_q)$ denote neighboring sets of $v_u$ and $v_q$, respectively. After obtaining the final communities, we define the edge weight sum (EWS) of each node $v_i$ in each community as $EWS_{i} = \Sigma^\infty_{h1=1}\Sigma^n_{j=1}[\alpha^{h1}*k_{j}]*(A^{h1})_{ij}$, where $(A^{h1})_{ij}$ is the number of paths from node $v_i$ to node $v_j$ of length $h1$, $k_j$ denotes the degree of node $v_j$ and $\alpha$ is a decay factor (we choose $\alpha = 0.1$). The nodes with the largest value of EWS are chosen as the seeds in each layer. To obtain $K$ seed nodes, the number of seeds chosen from each community is positively correlated with the size of the community.


\textbf{KSN}~\cite{kuhnle2018multiplex} is a knapsack seeding algorithm that first uses the CELF++ algorithm~\cite{goyal2011celf++} to obtain the expected influences of seed nodes of different sizes. The optimal seeds are determined by combining the solutions of each layer such that the sum of the expected influence spread of the selected seed sets is maximized and the size of the union of the selected seed sets is equal to $K$.

\textbf{CIM}~\cite{katukuri2022cim} is a clique-based heuristic algorithm that identifies seed nodes based on maximal clique\footnote{In a network, if a complete graph $C$ is not contained in any other complete graph then $C$ forms a maximal clique.}. CIM finds all maximal cliques in the multilayer network and ranks these cliques in descending order according to their size. Then CIM selects nodes with the highest degree in each maximal clique as seeds. If the number of maximal cliques $c < K$, the algorithm continues to select nodes with second largest degree from the maximal cliques following the descending order of their size. The selection process continues until the size of the seed node set reaches $K$.

\textbf{LAD} measures the importance of a node by calculating the layered average degree (LAD) for each node, i.e., $(\Sigma^L_{l=1}k^l_{i})/L$, $k^l_{i}$ is the degree of node $v_i$ in the $l$-th layer, $L$ is the number of layers. LAD algorithm selects $K$ nodes with the highest LAD values in the network as seed nodes.

\textbf{CD} is a heuristic algorithm based on coupling degree. We calculate the coupling degree $D_v$ of each node $v$ and pick the $K$ nodes with the highest coupling degrees to be the seed nodes.

\textbf{Eigenvector} considers a node's importance based on the importance of its neighboring nodes. The eigenvector first calculates the eigenvector centrality values of all nodes in each layer. The final eigenvector of each node is the average of its centrality values of eigenvectors from each layer. We choose the $K$ nodes with the highest eigenvector centrality values as seed nodes.

\textbf{Pagerank} first computes the PageRank values of each node in each layer. The final PageRank value of a node is the average of its PageRank values from each layer. The $K$ nodes with the highest final PageRank values are chosen as seeds.

\textbf{Betweenness} centrality first computes the betweenness of a node in each layer. The final betweenness value of a node is the average of the betweenness values over all the layers. The $K$ seed nodes with the highest Betweenness values are selected as seeds. 

The experiments are conducted to compare the effectiveness and robustness of different algorithms. Using the MIC model, we count the number of activated nodes at the end of the spreading process as the spread influence of a set of seed nodes. In our experiments, the size of the seed set $S$ ranged from 1 to 50 and the spread influence is the average over 1000 Monte Carlo simulations. All algorithms were implemented in Python and executed independently on a server with a 2.20GHz Intel(R) Xeon(R) Silver 4114 CPU and 90GB of memory.

\section{Performance evaluation}
\label{Results}
\subsection{Experimental results on synthetic multilayer networks}
We evaluate the performance of our algorithm on synthetic multilayer networks created using the ER (Erdős-Rényi), BA (Barabási-Albert), and WS (Watts-Strogatz) models, with network size $N=1500$. Specifically, an ER network is generated by starting with an empty network containing $N=1500$ nodes and traversing all pairs of nodes $(v_i, v_j) (v_i\neq v_j)$, the edge between node $v_i$ and node $v_j$ is added with a probability of $0.002$.  To construct a WS network, the process begins with a cyclic network of 1500 nodes. Each node is connected to its four closest neighbors. Then, each edge is randomly reconnected with a probability of 0.3, introducing randomness into the network. A BA network is established with an initial network that has three nodes and two edges. Whenever a new node is added to the network, it will be connected to two existing nodes through a preferential attachment mechanism. This means that nodes tend to link up with nodes that have a higher degree. Given the three networks obtained above, we combine each two of them to form two-layered networks, namely ER-ER, BA-BA, WS-WS, ER-BA, ER-WS, and BA-WS.

\begin{figure*}[!ht]
\centering
	\includegraphics[width=\linewidth]{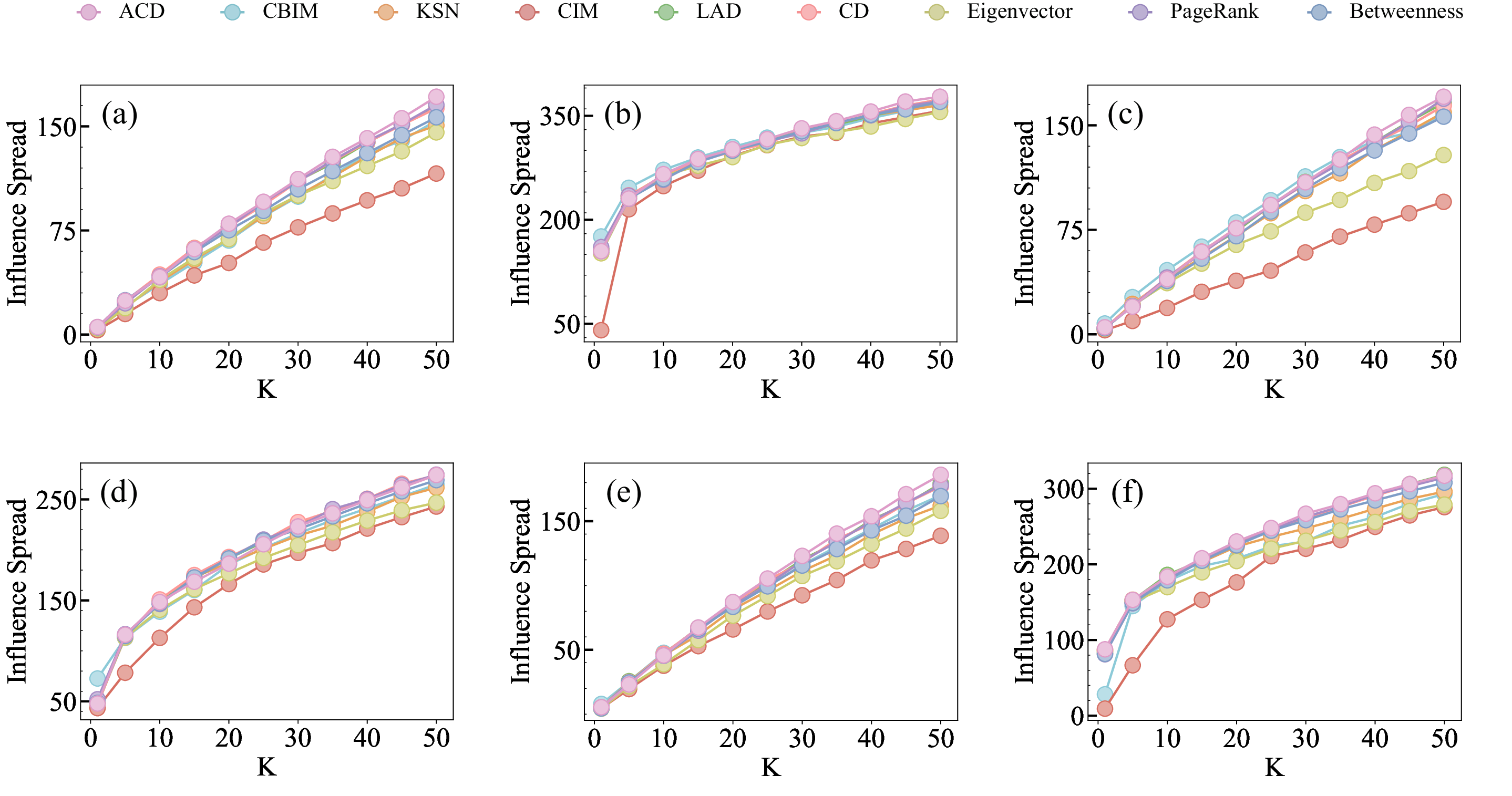}
 \caption{Spreading influence of different algorithms with different seed node sizes in artificial two-layered networks: (a) ER-ER; (b) BA-BA; (c) WS-WS; (d) ER-BA; (e) ER-WS; (f)BA-WS. We show the results for $K$ ranges from 1 to 50 and spread probability $p$=0.1.
 }
 \label{fig: artificial networks}
\end{figure*}

Furthermore, we conduct the MIC spreading model on the synthetic multilayer networks with spreading probability $p = 0.1$ and use influence maximization algorithms to identify seeds. The results in Figure~\ref{fig: artificial networks} demonstrate that the proposed ACD algorithm surpasses the other algorithms in terms of spreading capacity. On the other hand, the heuristic algorithms that disregard the interconnection between layers perform poorly, sometimes even worse than the centrality-based methods.

\newpage
\clearpage
\subsection{Experimental results on real-world multilayer networks}

\begin{figure*}[!ht]
\centering
	\includegraphics[width=\linewidth]{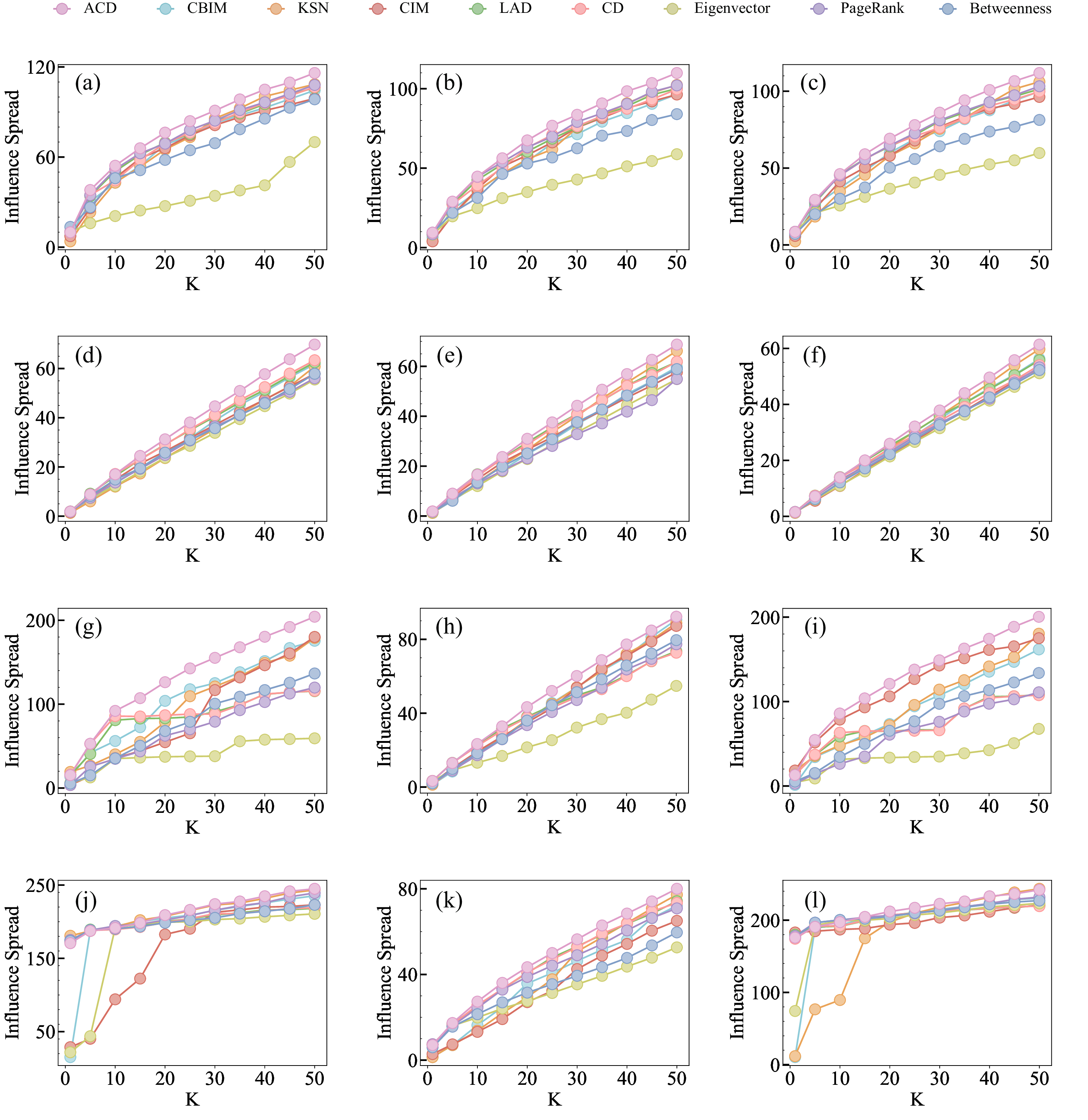}
 \caption{Influence spread of different algorithms with different seed node sizes in two-layered real-world networks: (a) Adv-Dis; (b) Adv-Fri; (c) Dis-Fri; (d) U-O; (e) U-D;(f) O-D;(g) MP-QBM; (h) MP-QB; (i) QBM-QB; (j) Phy-AI; (k) Phy-SI; (l) SI-AI.  We set $p$=0.1.}
 \label{fig: seed_influence}
\end{figure*}

\begin{figure*}[!ht]
\centering
	\includegraphics[width=\linewidth]{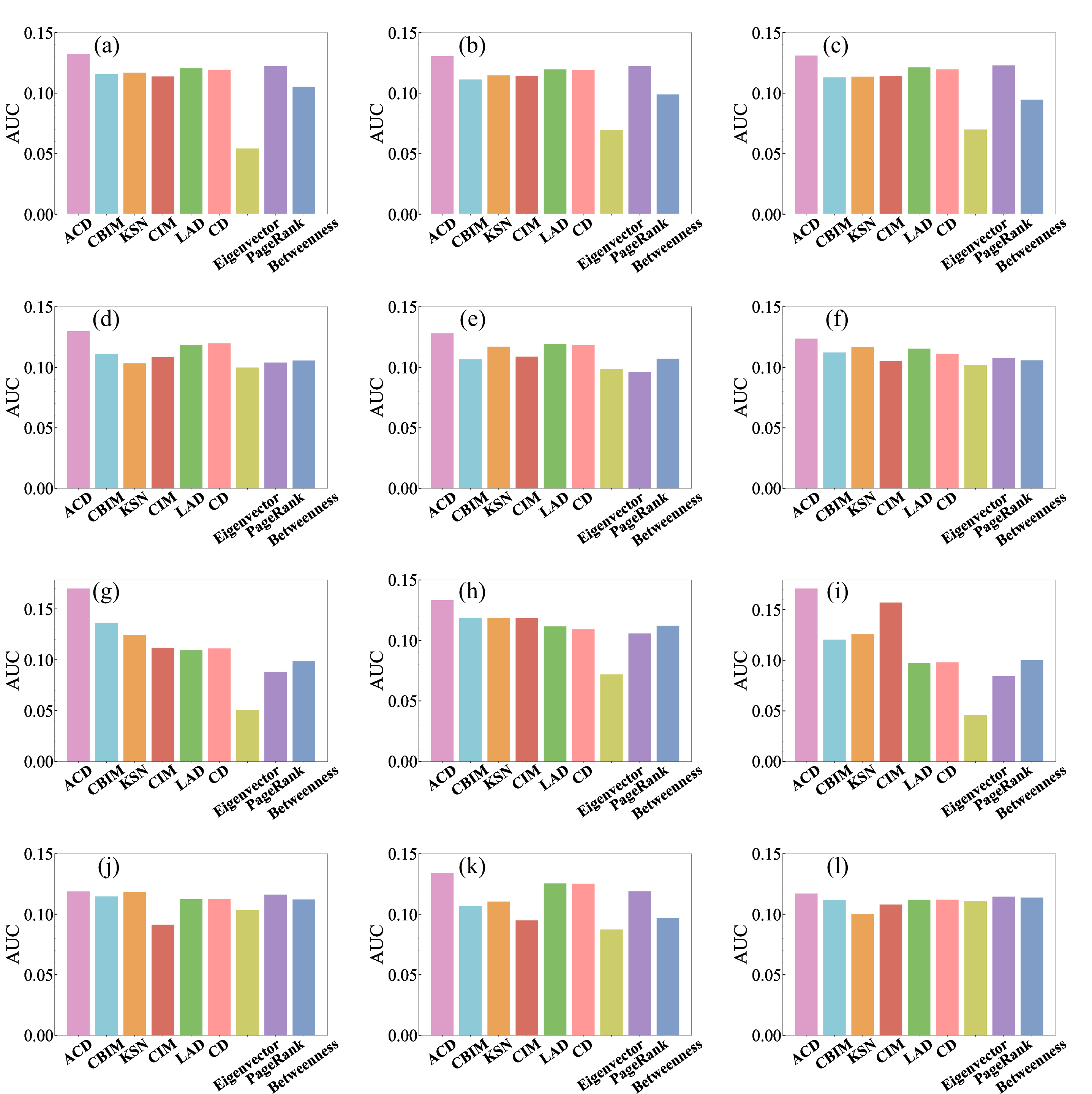}
 \caption{Histogram of AUC scores for different algorithms in Figure~\ref{fig: seed_influence}: (a) Adv-Dis; (b) Adv-Fri; (c) Dis-Fri; (d) U-O; (e) U-D;(f) O-D;(g) MP-QBM; (h) MP-QB; (i) QBM-QB; (j) Phy-AI; (k) Phy-SI; (l) SI-AI. We set $p$=0.1.}
 \label{fig: auc}
\end{figure*}

We further evaluate the performance of ACD on multilayer networks generated from four datasets, CKM, Transport, Arxiv, and C.elegans, using the MIC spreading model described in the Section of Preliminary definition. In the MIC model, the contagion probability is set to $0.1$. We use the final number of infected, i.e., influence spread in the vertical coordinate shown in Figure~\ref{fig: seed_influence}, deduced by the seeds as a metric to evaluate the performance of ACD. The average performance of each algorithm is represented by the area under the influence spread curve, as seen in Figure~\ref{fig: auc}.  Combining the two figures, we observe that ACD performs the best across all multilayer networks. Remarkably, ACD, LAD, and CD are all degree-based methods, the gap between ACD and the other two indicates that selecting nodes with low influence overlap can dramatically improve the accuracy of seed node identification. Our algorithm is one of several heuristic approaches to address the influence maximization problem, which is the same as CBIM and CIM. KSN is a greedy algorithm that sacrifices speed for improved performance. Nevertheless, the three algorithms are not as effective as ACD, which could be due to their disregard for the coupling effect between layers when constructing the algorithms.
Eigenvector, PageRank, and Betweenness, which are global centrality measures, do not work well on empirical multilayer networks, particularly Eigenvector. This could be due to the presence of isolated nodes in multilayer networks. The eigenvector is not ideal since isolated nodes lack connections from other nodes to affect their eigenvector values, leading to an overestimation of their importance.

\newpage
\clearpage
The time cost of different algorithms is presented in Table~\ref{time cost} when the seed size is $K=50$. We remark that the time cost is only for the seed node selection process without considering the MIC contagion process. 
ACD requires less time to complete than CBIM, KSN, and Betweenness. The difference could be attributed to additional operations, such as community detection and merging in CBIM as well as the greedy strategy in KSN for each layer of the network. Furthermore, for the betweenness centrality, it takes a lot of time to compute the shortest paths between each pair of nodes in the network.
 ACD is an iterative algorithm that considers both the influence of a node as well as the influence overlap between nodes, resulting in a slightly higher time cost compared to CIM, CD, Degree, Eigenvector, and PageRank. Overall speaking, ACD is more effective than baselines and requires an acceptable amount of time to identify seed nodes for IM in multilayer networks.

\begin{figure*}[!ht]
\centering
	\includegraphics[width=\linewidth]{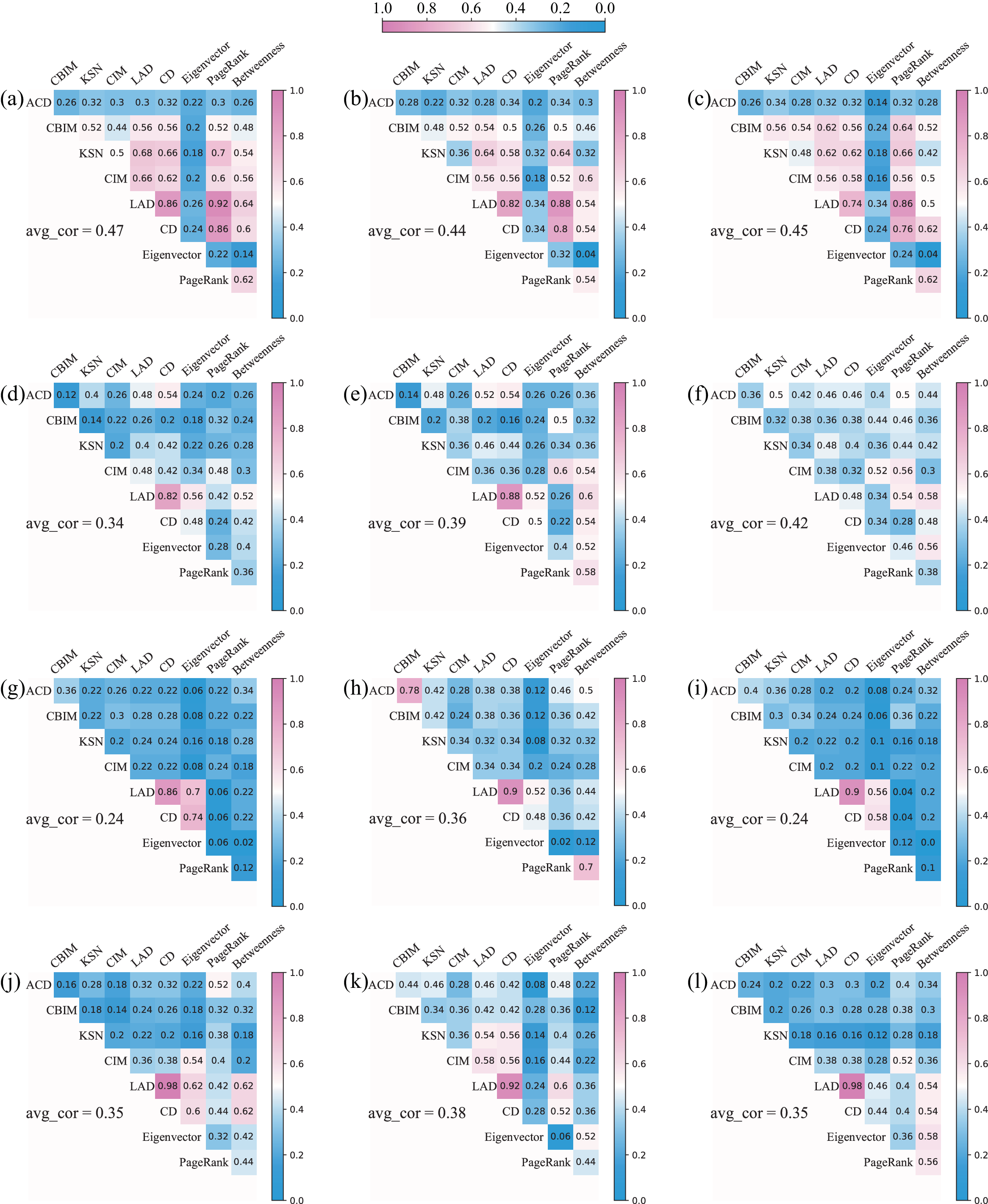}
 \caption{Correlation analysis of seed sets: 
 (a) Adv-Dis; (b) Adv-Fri; (c) Dis-Fri; (d) U-O; (e) U-D;(f) O-D;(g) MP-QBM; (h) MP-QB; (i) QBM-QB; (j) Phy-AI; (k) Phy-SI; (l) SI-AI. The avg\_cor is given by the mean of all values in each of the upper triangular matrices. We set $K$=50.}
 \label{fig: seed_overlap}
\end{figure*}

We show the correlation between ACD and the other baselines by computing the seed similarity between them for the seed size $K=50$. Specifically, for two algorithms $\mathcal{A}$ and $\mathcal{B}$, we suppose that the seed sets obtained by them are $S_{\mathcal{A}}$ and $S_{\mathcal{B}}$. Therefore, the similarity of the seed sets is defined as $\frac{|S_{\mathcal{A}}\cap S_{\mathcal{B}}|}{K}$. The illustration in Figure~\ref{fig: seed_overlap} displays the similarity between the seeds acquired by distinct algorithms in various multilayer networks. In addition, the average similarity value ($avg\_cor$) is presented, which is the mean of all values in each of the upper triangular matrices. The results show that the seeds obtained by ACD are quite different from the baselines, with most of the similarity values lower than the average values. 
\begin{table*}[!ht]
	\centering
	\caption{\label{time cost}\small Time cost for different algorithms.
 The running time is focused on the seed node selection process without considering the MIC contagion process. We set $K$=50.}
	\label{tab:1} 
    \resizebox{0.7\linewidth}{!}{
	\begin{tabular}{ccccccc cccc}
		\hline\hline\noalign{\smallskip}	
		Dataset& multiplex & ACD  & CBIM  & KSN &CIM & LAD &CD &Eigenvector &PageRank &Betweenness\\
  \noalign{\smallskip}\hline\noalign{\smallskip}
        & Ad-Dis  &0.0038  &9.7406 &33.1708  &0.0070   &0.0004 &0.0021 &0.3403  &0.0026  &0.0414 \\
    CKM & Ad-Fri &0.0038  &9.4953 &30.4458  &0.0071  &0.0003 &0.0023 &0.3452  &0.0027  &0.0405 \\
          & Dis-Fri  &0.0036 &10.2109 &29.6869  &0.0069  &0.0003 &0.0018 &0.0665  &0.0030   &0.0430 \\  
		\noalign{\smallskip}\hline\noalign{\smallskip}
		    & U-O &0.0050   &0.3269 &21.2453  &0.0044  &0.0010  &0.0016 &0.0153  &0.0031  &0.0764  \\
	Transport &U-D &0.0051  &0.3036 &17.8422  &0.0043  &0.0010 &0.0011  &0.0125  &0.0030   &0.0700\\
              & O-D  &0.0047  &0.0382 &18.1863  &0.0032  &0.0012 &0.0006 &0.0072  &0.0028  &0.0086\\
         \noalign{\smallskip}\hline\noalign{\smallskip}
         & MP-QBM &0.1971  &16.4488 &406.7342   &0.0334   &0.0610 &0.0182   &0.0725   &0.0074   &1.3757\\
        Arxiv & MP-QB   &0.1859   &2.4575 &679.5698  & 0.0110   & 0.0590 &0.0053   &0.0396   &0.0040    &0.0463\\
          & QBM-QB  &0.2004   &17.4843 &1699.3893    &0.0295   & 0.0564 &0.0189   &0.0794    &0.0080 &1.3807\\
        \noalign{\smallskip}\hline\noalign{\smallskip}
         & Phy-AI &0.0523 & 563.6921 &1345.1427   & 0.0172   & 0.0154  &0.0073  &0.0358   & 0.0173 &1.4071\\
    Celegans  & Phy-SI &0.0508   &1.1359 & 266.7521 &  0.0053 &  0.0159 &0.0013  &0.0089   &0.0041   &0.0162\\
          & SI-AI  &0.0516  &547.2645 &1283.2126    &0.0133    &0.0155 &0.0062   &0.0325    &0.0060 &1.3649\\                                        
		\noalign{\smallskip}\hline
	\end{tabular}
    }
\end{table*}

\clearpage
\subsection*{Robustness analysis}
\begin{figure*}[!ht]
\centering
	\includegraphics[width=\linewidth]{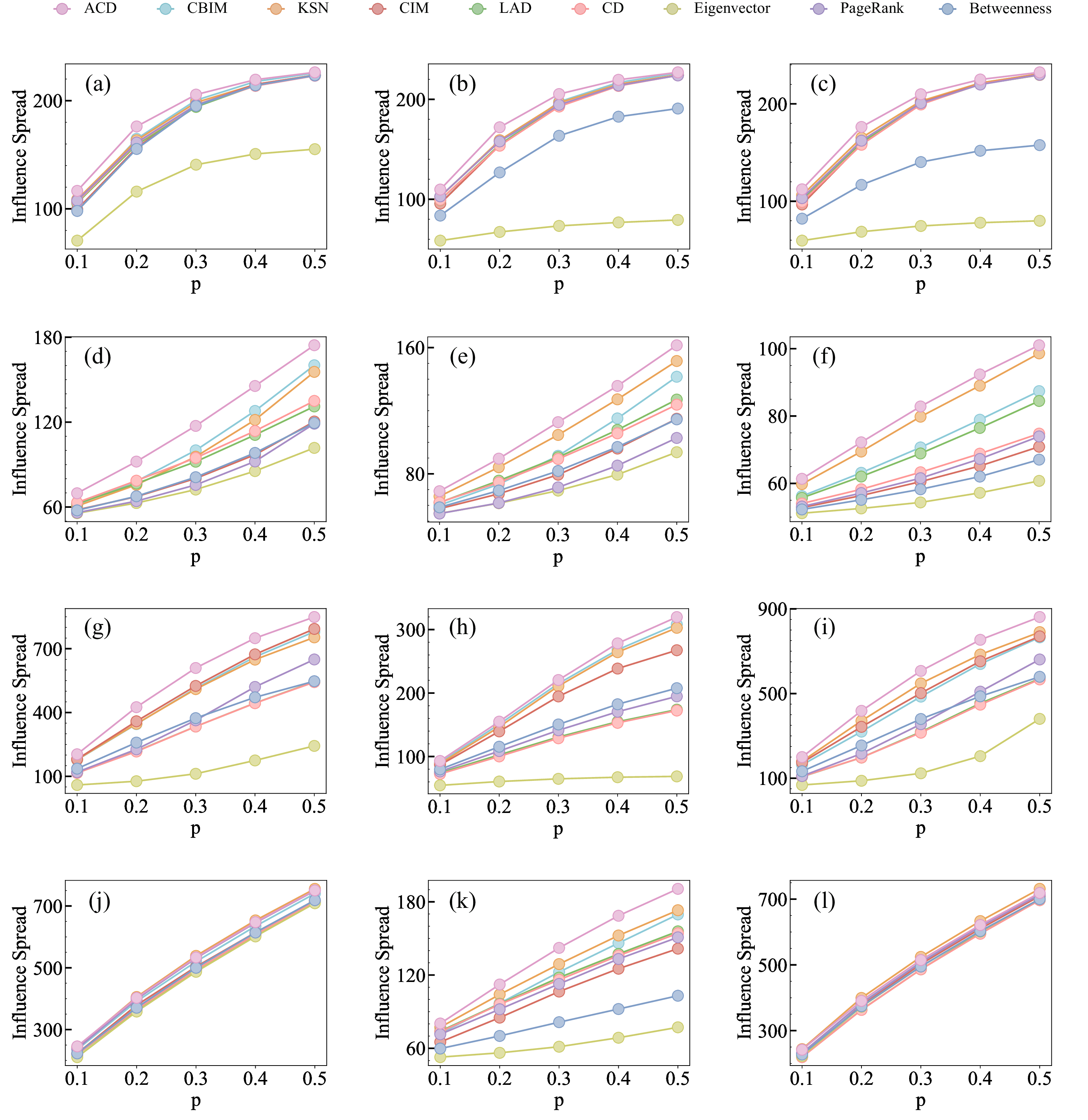}
 \caption{Spreading influence of different algorithms with different propagation probability p in two-layered real-world networks: 
 (a) Adv-Dis; (b) Adv-Fri; (c) Dis-Fri; (d) U-O; (e) U-D;(f) O-D;(g) MP-QBM; (h) MP-QB; (i) QBM-QB; (j) Phy-AI; (k) Phy-SI; (l) SI-AI. We set $K$=50.}
 \label{fig: P}
\end{figure*}

The robustness of ACD is evaluated by changing the contagion probability $p$ of MIC when the seed size is $K=50$, as shown in Figure~\ref{fig: P}. We present the results of $p$ ranging from $0$ to $0.5$ with an interval of $0.1$, as a value of $p$ that is too high could cause the propagation process to cover most nodes in the network regardless of the selected seeds. As $p$ increases, the spread of influence also increases,  yet ACD usually performs best across different networks. Figure ~\ref{fig: P} j and l demonstrate two exceptions, where KSN has the highest performance when the spread probability is greater than $0.3$, while ACD is still the second best. The results demonstrate the efficacy of the ACD algorithm in a variety of spread probabilities, particularly in terms of its superior performance in most cases.

\begin{figure*}[!ht]
\centering
	\includegraphics[width=\linewidth]{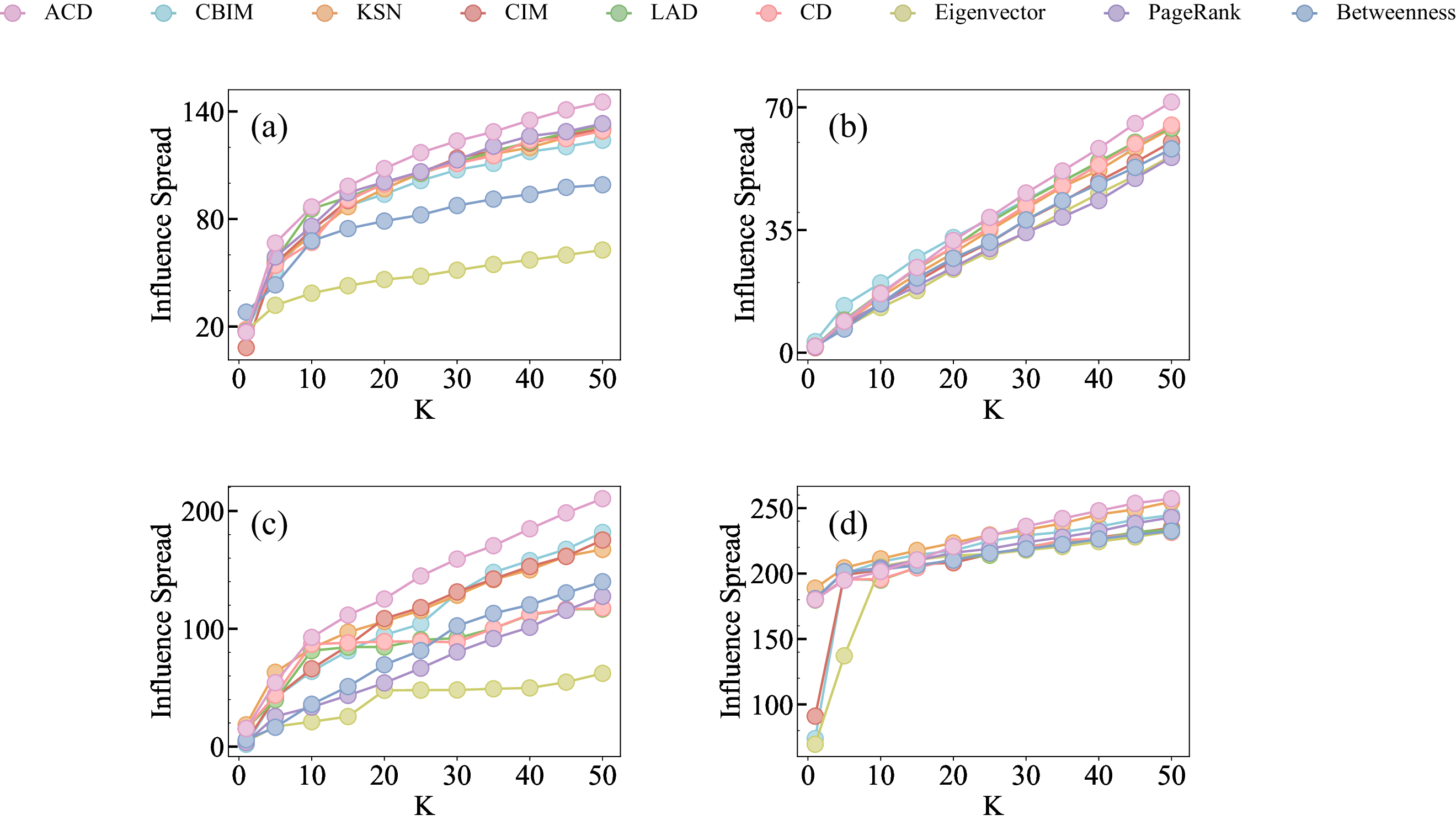}
 \caption{Spreading influence of different algorithms in three-layered real-world networks: 
 (a) Adv-Dis-Fri; (b) U-O-D; (c) MP-QBM-QB; (d) Phy-AI-SI. We set $p$=0.1.}
 \label{fig: three_multilayer}
\end{figure*}

We further test the effectiveness of ACD in three-layer networks when the contagion probability $p=0.1$. As shown in Figure~\ref{fig: three_multilayer}, ACD performs almost optimally in all four three-layer networks, with some exceptions when $K$ is small. We observe that ACD performs better when $K$ is larger. The good performance of ACD in two- and three-layer empirical networks indicates its robustness across multilayer networks from different areas.

\clearpage
\section*{Conclusion}
The main challenge of solving the IM problem on a multilayer network is how to combine topological information of one node upon different layers, especially incorporating the coupling effect between layers. In this work, we propose a method named Adaptive Coupling Degree (ACD), which iteratively selects seed nodes with low influence overlap between each other and a high coupling degree. Additionally, a spreading model, i.e., MIC, which considers a piece of information spreading through a multilayer network and enhances inter-layer spread interactions, is proposed to quantify the spread influence of the nodes selected by our method.

To validate the effectiveness of our algorithm, we conduct extensive experiments on four empirical multilayer networks, which are generated by complex systems from different scenarios, such as social networks, transportation networks, and biological networks. 
We compare the performance of ACD with eight baseline algorithms, and the experimental results demonstrate that ACD exhibits a significantly wider influence spread than the baseline algorithms. Moreover, we assess the robustness of ACD across various empirical and synthetical multilayer networks, highlighting its effectiveness in diverse network structures.

The limitation of our work is that we only consider the fundamental topological structure to design the algorithm for the IM problem. However, researchers have claimed that the dynamic process may also affect the effectiveness of the algorithms~\cite{gong2016influence, lu2016big}. In future work, the dynamic aspect of the nodes should also be considered and a general framework that can be adapted to different spreading models, such as as multilayer linear threshold models~\cite{zhong2021influence}, multilayer susceptible-infected-susceptible (SIS) models~\cite{pare2022multilayer}, and other propagation mechanisms~\cite{Br2020}, should be designed. Additionally, more effort should be made to solve the extensive IM problems on multilayer networks, i.e., budgeted IM~\cite{nguyen2013budgeted,banerjee2019combim}, competitive IM~\cite{bharathi2007competitive,bozorgi2017community}, and targeted IM~\cite{cai2020target,calio2021attribute,liang2023targeted}.

\section{Declaration of competing interest}
The authors declare that they have no known competing financial interests or personal relationships that could have appeared to influence the work reported in this paper.
\section{Data availability}
Data will be available on request.
\section{Acknowledgement}
This work was supported by the Natural Science Foundation of Zhejiang Province (Grant No. LQ22F030008), the Natural Science Foundation of China (Grant No. 61873080) and the Scientific Research Foundation for Scholars of HZNU (2021QDL030).

\bibliographystyle{elsarticle-num}
\bibliography{TempExample}

\end{document}